\documentclass{article}
\usepackage[margin=1in]{geometry}
\usepackage[T1]{fontenc}
\usepackage{epsfig}
\usepackage{amssymb}
\usepackage{amsmath}
\usepackage{amsfonts}
\usepackage{pgf-pie}
\usepackage{booktabs}
\usepackage{blindtext}
\usepackage{alltt}
\usepackage{hyperref}
\usepackage{listings}
\usepackage{multirow}
\usepackage{array}
\usepackage{color}
\usepackage{apacite}
\definecolor{gray}{rgb}{0.4,0.4,0.4}
\definecolor{darkblue}{rgb}{0.0,0.0,0.6}
\definecolor{cyan}{rgb}{0.0,0.6,0.6}

\lstdefinelanguage{XML}
{
  morestring=[b]",
  morestring=[s]{>}{<},
  morecomment=[s]{<?}{?>},
  stringstyle=\color{black},
  identifierstyle=\color{darkblue},
  keywordstyle=\color{cyan}
}

\title{Scientific Table Search Using Keyword Queries}
\author{
  Kyle Yingkai Gao\\
    Language Technologies Institute\\
    Carnegie Mellon University\\
    Pittsburgh, PA 15213, USA\\
  \texttt{yingkaig@cs.cmu.edu}
  \and
  Jamie Callan\\
    Language Technologies Institute\\
    Carnegie Mellon University\\
    Pittsburgh, PA 15213, USA\\
  \texttt{callan@cs.cmu.edu}
}

\begin{document}

\maketitle

\begin{abstract}
Tables are common and important in scientific documents, yet most text-based document search systems do not capture structures and semantics specific to tables.  How to bridge different types of mismatch between keywords queries and scientific tables and what influences ranking quality needs to be carefully investigated.  This paper considers the structure of tables and gives different emphasis to table components.  On the query side, thanks to external knowledge such as knowledge bases and ontologies, key concepts are extracted and used to build structured queries, and target quantity types are identified and used to expand original queries.  A probabilistic framework is proposed to incorporate structural and semantic information from both query and table sides.  We also construct and release TableArXiv, a high quality dataset with 105 queries and corresponding relevance judgements for scientific table search.  Experiments demonstrate significantly higher accuracy overall and at the top of the rankings than several baseline methods.
\end{abstract}

\section{Introduction}
\label{sec:intro}
Search of scientific publications has become a routine task due to search services such as Google Scholar\footnote{\url{https://scholar.google.com/}}, Microsoft Academic Search\footnote{\url{http://academic.research.microsoft.com/}}, CiteSeerX\footnote{\url{http://citeseerx.ist.psu.edu/index}}, Semantic Scholar\footnote{\url{https://www.semanticscholar.org/}}, and the ACM Digital Library\footnote{\url{http://dl.acm.org/}}.  These services use full-text ranking to find papers (documents) that match the query.  They are useful for discovering publications about a topic, which one may read at one's leisure.  However, if one's goal is to find a specific value such as the mass of a meson or to find values reported recently for the effectiveness of a particular algorithm on a particular dataset, these services are less useful because the desired information is likely to be in a table.  A person must read the top-ranked papers to find desired information, which is time-consuming.  The ranking of papers may also be less accurate for this task because typically tabular data is just a footnotesize part of the bag of words that represents the document.

In these situations, it is more useful for the search engine to return tables of data, not scientific papers that contains tables of data.  Table retrieval is similar to document retrieval insofar as the system must produce a ranking of objects; however, there are also important differences.  The structure of a table is essential to understanding its meaning, however typical bag-of-words retrieval methods discard that structure.  The document that contains the table provides important contextual, descriptive, and clarifying information about the table contents, however much of the document text may also be unrelated to the table.  If a table search engine is to support keyword queries effectively, it must have strategies for addressing these types of mismatch.

Prior research often addressed table retrieval in the context of table recognition and extraction, however they are separate problems.  \textit{Table recognition and extraction} is the problem of recognizing and extracting tabular content in document formats that have unreliable table markup (e.g., HTML) or no explicit table markup (e.g., PDF).  \textit{Table retrieval} is the problem of ranking tables by how well they satisfy a query.  The problems are often intermingled because the web is an important source of tabular data.  Recognition and extraction from HTML and PDF are far from perfect, thus a table retrieval system for these sources must cope with data of widely varying quality.  Zanibbi et.\ al.\cite{ZanibbiBC04} provide a comprehensive survey of this line of research.  

Publishers such as arXiv.org\footnote{\url{http://arxiv.org/}}, Elsevier\footnote{\url{https://www.elsevier.com/}}, the American Institute of Physics\footnote{\url{https://www.aip.org/}}, the Chemical Abstracts Service\footnote{\url{https://www.cas.org/}}, and SAE International\footnote{\url{https://www.sae.org/}} have large digital libraries of scientific publications that contain tabular data.  Often these are the same papers that can be scraped from the web in PDF format.  However, unlike data scraped from the web, papers from these digital libraries are available in XML, LaTeX, or other `source code' formats that make table recognition and extraction simple problems and that produce data with few errors.  This type of data makes it possible to focus entirely on ranking tables accurately.  Knowing the publisher and paper attributes also enables the use domain knowledge or other resources that may improve understanding of the query or table.

This paper focuses on table retrieval as it would be performed by scientific publishers that have the `source code' for each document.  The problems of table recognition and extraction from PDF and other `compiled' formats, although important, are beyond the scope of this paper.  We assume that typical unstructured keyword queries are used to retrieve tables from scientific publications.  Each table is treated as a retrieval unit.  

The task is challenging because there are several types of mismatch between a query and a table.  Authors tend to use shortened names and symbols in tables whereas query writers are likely to use more complete names (\textit{textual mismatch}).  Typical keyword queries are unstructured, but tables have a visual and relational structure that is important to understanding their meaning (\textit{structural mismatch}).  Tabular data is likely to be associated with units that implicitly indicate specific quantity types, but query writers may not explicitly specify the types of quantities that they desire (\textit{semantic mismatch}).  Figure~\ref{fig:challenge} illustrates the challenges.  The table does not explicitly mention x-ray emission, but its row headers indicate that it is related to x-ray emission instead of other information about x-rays (\textit{textual mismatch}).  The query is unstructured, but the desired concepts appear in different parts of the table (e.g. \texttt{x-ray} in the article title), organized in a way that signals a relationship between them (\textit{structural mismatch}).  Finally, the query does not indicate the type of value desired, while the table uses kiloelectron-volts (keV) to specify the value type (\textit{semantic mismatch}).

This paper presents MaitreD, a probabilistic ranking framework for scientific table retrieval using keyword queries (``it gets you the table you want'').  
MaitreD  extracts key concepts from keyword queries, identifies quantity types, and uses unit symbols to transform unstructured bag of words queries into more effective structured queries.  Tables are represented as structured objects that have multiple elements or representations.  Different parts of the query are matched with differing weights to different parts of the table to make more effective use of all of the evidence available.
% Together with the original query, all evidences are matched with different emphasis of table components to generate ranking probabilistically.  
Experiments demonstrate that these methods significantly improve retrieval of scientific tables compared to strong, state-of-art baseline methods.

This paper also presents TableArXiv, a new dataset with 631,734 publicly-available documents, 105 queries, and corresponding relevance judgements specific to scientific table retrieval that others can use to reproduce and extend the research reported here.\footnote{\url{http://www.cs.cmu.edu/~callan/Data/}}

\begin{figure}[t]
    \centering
    \begin{tabular}{l|c}
    \multicolumn{2}{l}{\textbf{Query:} \texttt{x-ray emission spectra}} \\
    \multicolumn{2}{c}{} \\ 
    \multicolumn{2}{p{8cm}}{\textbf{Matching Table:}} \\
    \multicolumn{2}{p{8cm}}{\textbf{Article Title:} \texttt{Broad band X-ray spectroscopy of A0535+262 with SUZAKU}} \\ 
    \multicolumn{2}{p{8cm}}{\textbf{Caption:} \texttt{Best-fit parameters of the phase-averaged spectra for A0535+262 during 2005 observation with $1\sigma$ errors}} \\
    \multicolumn{2}{p{8cm}}{} \\ \hline 
    Parameter & Value \\ \hline
    $kT_{BB}$ (keV) & $0.16 \pm 0.01$ \\
    Iron line width (keV) & $0.03 \pm 0.02$ \\
    Iron line flux$^a$ & $0.8 \pm 0.1$ \\
    CYAB width (W, keV) & $9.5 \pm 1.2$ \\ \hline
    \end{tabular}
    \caption{A query and table that illustrate the challenges of table search in scientific domains.}
    \label{fig:challenge}
\end{figure}

The contributions of this research are as follows:
\begin{itemize}
\item It uses external knowledge to automatically identify key concepts and quantity types in keyword queries and uses them to structure and expand queries;
\item It defines a probabilistic model that flexibly incorporates structural and semantic information from queries and tables;
\item It describes a high quality, publicly available dataset for scientific table retrieval; and
\item Experiments show that MaitreD is significantly more effective at table retrieval than several state-of-art document-ranking and table-ranking algorithms.
\end{itemize}

The rest of this paper is organized as follows.  Section~\ref{sec:relate} reviews research that is related to the work presented here.  Sections~\ref{sec:table} and~\ref{sec:query} describe how to represent tables and queries.  The representations are used by the probabilistic ranking model presented in Section~\ref{sec:model}.  Section~\ref{sec:data} presents and reviews the construction of the TableArXiv dataset.  Finally experimental results followed by discussion and conclusions are described in Sections~\ref{sec:eval} and ~\ref{sec:conclusion}.

\section{Related Work}
\label{sec:relate}

% Table Extraction

Research on table recognition and extraction predates most of the research on table retrieval.  As the dominant document format changed over time, approaches were developed based on visual clues \cite{cesarini2002trainable,kieninger1998table,Rus:1994:UWS:866729}, pure text documents \cite{DBLP:conf/acl/NgLK99,Pyreddy1997DL,Wei2006InfRetr}, HTML documents \cite{DBLP:conf/coling/ChenTT00,DBLP:conf/www/CrestanP10,DBLP:conf/widm/YangL02}, and PDF documents \cite{DBLP:conf/aaai/FangMTG12,DBLP:conf/jcdl/LiuBMG07,DBLP:conf/cikm/LiuMG08}.  As with many other research fields, mainstream research migrated from heuristics \cite{DBLP:conf/coling/ChenTT00,kieninger1998table,DBLP:conf/jcdl/LiuBMG07,Pyreddy1997DL} to machine learning methods that use a variety of features \cite{DBLP:conf/www/CrestanP10,DBLP:conf/aaai/FangMTG12,DBLP:conf/cikm/LiuMG08,DBLP:conf/acl/NgLK99,Wei2006InfRetr}.

% Table Information Extraction

As a downstream application of table recognition, there also has been substantial interest in information extraction from tables, especially HTML tables in web pages.  Much of the research focuses on discovering~~\cite{Yin2011WWW}, identifying~~\cite{Venetis2011PVLDB}, and augmenting~~\cite{Yakout2012SIGMOD, Zhang2013SIGMOD} entities and entity-attribute relations from tables, for example, learning names of football teams~\cite{DBLP:conf/emnlp/WangC09,DBLP:conf/wsdm/DalviCC12}.  Venetis et\ al.~\cite{Venetis2011PVLDB} demonstrates that augmenting entity descriptions from tables can help search tasks.  WebTables~\cite{Cafarella2008PVLDB} shows that using features of both keyword matching and schema coherency is effective for table search.

% Table Search

% Table fields importance - structure mismatch
Table retrieval is not a new problem, although it received less attention until recently.
Compared to most forms of text retrieval, tables have an impoverished text representation.  They are short, and their content often has a high proportion of abbreviations, special symbols, and/or numbers that are not likely to occur in queries.  In scientific tables, the situation is even worse since they usually contain a large portion of numbers and less text.  Thus, it has become common to represent a table by its contents and also its surrounding document context, treating the different sources of information as different fields in a document.  TinTin~\cite{Pyreddy1997DL} was one of the first systems to consider the differing importance of table parts to improve ranking.  However, which fields of a table to match and how much weight to give each field were manually determined, thus requiring a skilled and knowledgeable user.  Later, the importance of each field was determined either by experience~\cite{Pyreddy1997DL,Wei2006InfRetr,Pimplikar2012CoRR} or by popular term distribution~\cite{Liu2007AAAI}.

% Table ranking

Ranking tables is often treated as similar to ranking documents that have hierarchical structure and sparse content.  TableRank~\cite{Liu2007AAAI} used a tailored TF-IDF weighting scheme and cosine similarity to rank tables.  Wei et\ al.~\cite{Wei2006InfRetr} used keyword question queries to rank table cells.  Thomas et al.~\cite{Thomas:2015:TSA:2838931.2838941} improved the ranking of table columns by inferring the data types of table columns and query terms.  WWT~\cite{Pimplikar2012CoRR} answered table queries by aggregating relevant columns retrieved and scored by a joint graphical model.  Recently Web-based table retrieval systems for HTML and PDF document formats have become more common, for example, Google Tables\footnote{\url{https://research.google.com/tables}} and Zanran\footnote{\url{http://zanran.com/}}, however their algorithms have not been published. 

% Semantic mismatch

The semantic mismatch between queries and table contents has sometimes been addressed by trying to understand numerical values in tables.  InforGather+~\cite{Zhang2013SIGMOD} labeled table columns with units, scales and timestamps to make values comparable.  For the same purpose, QEWT~\cite{Sarawagi2014KDD} recognized units in tables with both high precision and recall using a probabilistic context-free grammar (PCFG).  

Due to the complex structure and content type in tables, recovering semantics from tables is still difficult and prone to noise.  In this paper we propose to bridge the mismatch on the query side, using query expansion and query structure understanding techniques.

% Query expansion using knowledge base, ontology, or controlled vocabulary

There is a long history of research on using query expansion to improve retrieval accuracy.  Pseudo-relevance feedback is often used for document retrieval, however it is not used often for table retrieval, perhaps due to the impoverished text representations found in most table retrieval systems.  Query expansion can also be done using information external to the corpus, such as thesauri~\cite{KekalainenJ98}, knowledge bases, ontologies, and controlled vocabularies.
Wikipedia~\cite{brandao2014learning,xu2009query,bendersky2012effective} and Freebase~\cite{Xiong2015ICTIR} are examples of general-purpose resources used for query expansion, and MeSH~\cite{lu2009empirical,lu2009evaluation} is an example of a domain-specific controlled vocabulary used for query expansion.  Recently Xiong et\ al.~\cite{Xiong2015CIKM} proposed a learning-based framework that incorporates both knowledge bases and controlled vocabularies into a single method of query expansion.
Regardless of the type of external information, these approaches share a high-level strategy that first links query terms to information (typically entities) in the external resource, and then follows those links to find additional information (often related terms) that is used to improve ranking accuracy.
% In section~\ref{sec:query_quantity} we will show that a classifier is used to link queries to quantity types and quantity units are used to expand our queries.

Most prior research on understanding query structure has been done for document retrieval, as opposed to table retrieval.  Noun phrases~\cite{CroftTL91} and named entity recognition~\cite{Callan:1995Trec} have been used to reformulate keyword queries into structured queries.  Heuristics, machine learning, and properties such as synonymy, centrality, and abstractness have been used to detect and reweight concepts in longer queries~\cite{Bendersky2008SIGIR,Bendersky:2010:LCI:1718487.1718492,LiWA09,DBLP:conf/cikm/ZhaoC10}.  This and much other similar research demonstrated that identifying and emphasizing key concepts in queries, especially long queries, can improve document ranking accuracy.  

In summary, query understanding and augmentation for table search has been less studied than enriched representations of table contents.  It is not yet clear how query reformulation techniques will influence table retrieval accuracy.  This paper investigates these problems, assuming keyword queries and a table-level granularity.

\section{Table Representation}
\label{sec:table}

In this work, tables are modeled as objects that have multiple content-based representations and an estimate of the table's quality.  The sections below describe how these representations and estimates are obtained.

\subsection{Table Contents}

Table retrieval is unlike document retrieval in that tables often contain few words that describe their contents; this is especially true for tables of numeric data.  Thus, it is useful to enrich the description of the table with information harvested from the document that contains it~\cite{Liu2007AAAI,Wei2006InfRetr}.  Our tables contain eight types of fields that describe their contents at three levels of detail.
\begin{itemize}
  \item \textbf{Document:} Information from the document provides a context for understanding the table.  The \texttt{title} and \texttt{abstract} fields are concise descriptions of the paper content.  (2 fields)
  \item \textbf{Table:} Table content is summarized by its \texttt{caption}, \texttt{referring sentence}, and \texttt{footnote} fields.  Captions and referring sentences are analogous to titles and in-link text for web pages -- concise, but highly informative.   (3 fields)
  \item \textbf{Cell:} The most detailed description of table contents is provided by \texttt{row header}, \texttt{column header}, and \texttt{cell value} fields.  These fields enable queries to match at a very fine granularity.  Although cell values might seem less useful because they are often specific numeric values, they may also contain quantity units that are useful for matching at the semantic level. (3 fields)
\end{itemize}
Figure~\ref{fig:fields} shows the multi-field  structure for the example table in Figure~\ref{fig:challenge} (abstract, footnote, and referring sentence added).  No single field determines whether a query matches the table.

Decomposing table-related text into multiple fields with different characteristics enables different strategies for matching query elements to table elements.

\subsection{Table Quality}
\label{sect:TableQuality}

Authority and quality estimates have become common in document retrieval.  Quality metrics are also useful for table retrieval.  Prior research considered metrics such as the coherence of the table schema, whether the table has column headers, and the length of table reference sentences, as well as document quality signals~\cite{Cafarella2008PVLDB,Liu2007AAAI}.

We propose the proportion of numeric content as an additional measure of table quality for science and engineering.  This measure is supported by two observations.  First, scientific results are usually in the form of numbers; text tables are often used to show examples (as in this paper).  Second, we observed that the people who created queries for the TableArXiv dataset were more interested in numeric values.  

This table quality feature is defined as:
\begin{equation*}
    p(t)=\frac{\sum_{\text{cell} \in t}\mathbb{I}(\text{\texttt{cell\_value} is numeric})}{\sum_{\text{\texttt{cell\_value}} \in t}\mathbb{I}(\text{\texttt{cell\_value}})}
\end{equation*}
where $\mathbb{I}(\text{\texttt{cell\_value} is numeric})$ is a indicator function that equals  1 when a cell is numeric and is zero otherwise.  A regular expression program determines whether a cell is numeric.  Typical numeric values include, but are not limited to, integers, floats, ranges, numbers with accuracies, and numbers followed by units.  The denominator counts the number of cells.  Experiments described in Section~\ref{sec:eval} investigate the effectiveness of this simple quality metric.

\begin{figure}[t]
    \centering
    \begin{lstlisting}[language=xml,breaklines=true,mathescape=true]
<table>
  <!------------ document-level -------------->
  <article-title>Broad band X-ray spectroscopy of A0535+262 with SUZAKU</article-title>
  <abstract>The transient X-ray binary pulsar A0535+262 was observed with Suzaku on 2005 September...</abstract>
  <!-------------- table-level --------------->
  <caption>Best-fit parameters of the phase-averaged spectra for A0535+262 during 2005 observation with 1sigma errors</caption>
  <referring-sentences>
    <sentence>The spectral parameters of the best-fit model obtained from the simultaneous spectral are given in Table 1</sentence>
  </referring-sentences>
  <footnotes>
    <footnote>a. flux estimated in 1.0-10.0 keV energy range.</footnote>
  </footnotes>
  <!-------------- cell-level ---------------->
  <column_headers>
    <column_header>Parameter</column_header>
    <column_header>Value</column_header>
  </column_headers>
  <row_headers>
    <row_header>kT_BB (keV)</row_header>
    <row_header>Iron line width (keV)</row_header>
    ...
    <row_header>CYAB width (W,keV)</row_header>
  </row_headers>
  <cell_values>
    <cell_value> 0.16$\pm$0.01 </cell_value> 
    ...
  </cell_values>
</table>
    \end{lstlisting}
    \caption{An XML representation of the multi-field index structure for the table in Figure~\ref{fig:challenge}. Text from different sources is stored in different fields.}    \label{fig:fields}
\end{figure}

\section{Query Representation}
\label{sec:query}

People find unstructured keyword queries a natural way to express their information needs.  Prior research shows that transforming unstructured queries into structured queries greatly improves ranking accuracy~(e.g., \cite{Bendersky2008SIGIR,bendersky2012effective,Callan:1995Trec,CroftTL91,KekalainenJ98,Metzler2007SIGIR}).  Our method of transforming unstructured scientific queries into structured scientific queries has two components:  Concept identification and weighting, and quantity expansion.

\subsection{Concept Identification and Weighting}
\label{sec:query_concept}
The first step in transforming an unstructured query to a structured query is to identify the most important concepts.  Two types of concepts were investigated:  entities and noun phrases.  Entities have the advantage that they have precise meanings, descriptions, characteristics, and relationships to other entities.  Noun phrases have the advantage of flexibility; they cover a much broader range of concepts than entities, including unanticipated concepts, and they do not require external resources (e.g., Wikipedia or another knowledge source).  Much prior research indicates that both types of concept can be useful in typical document ranking~\cite{Bendersky2008SIGIR,Callan:1995Trec,Xiong2015CIKM}, however their effectiveness for table retrieval has received less attention.

\textbf{Entity identification} (\textit{entity tagging}) was done by TagMe, a popular and effective system that annotates short texts with Wikipedia entities~\cite{Carmel2014SIGIR,Ferragina2012IEEESoftware}.  TagMe gives a score $\rho_i$ for each entity (concept) $c_i$ to indicate its confidence that the text string matches the entity identified.

TagMe's confidence score $\rho_i$ was normalized to produce a maximum likelihood estimate $p(c_i|q)$, the probability of concept $c_i$ given query $q$:
\begin{equation*}
    p(c_i|q) = \frac{\rho_i}{\sum_{c_j \epsilon q} \rho_j}
\end{equation*}
where the summation is over all concepts tagged in $q$. 

\textbf{Noun phrase identification} was done by MontyLingua~\cite{liu2004montylingua}, a commonly used end-to-end natural language tool.  MontyLingua does not provide a confidence score for each noun phrase annotation.  Instead, confidence scores were estimated by a method described by Bendersky and Croft~\cite{Bendersky2008SIGIR}.  A ranking-based classifier was trained to produce an estimate $h(c_i)$ that the noun phrase $c_i$ is a key concept.  The features were collection term frequency (ctf) and inverse document frequency (idf) scores for each type of table field except the \texttt{cell\_value} field type (7 types of table field, thus 14 features).
If a noun phrase was too long to have exact sequential matches in the corpus, we used the average ctf and idf of all of its consecutive two-term sub-sequences.  The estimated confidence score $h(c_i)$ was normalized by the sum of the noun phrase scores to estimate $p(c_i|q)$
\begin{equation*}
    p(c_i|q) = \frac{h(c_i)}{\sum_{c_j \epsilon q} h(c_j)}
\end{equation*}

\subsection{Quantity Expansion}
\label{sec:query_quantity}
Queries may use an abstract or general term to describe a desired quantity (e.g., \texttt{emission}), however tables are more likely to use specific units of measure (e.g., eV, or keV).  This mismatch can be overcome by normalizing table values to standard units, or by mapping abstract query terms to typical instantiations.  The former solution is common in manually-curated systems, but difficult to accomplish automatically.  QEWT~\cite{Sarawagi2014KDD} provides a PCFG based unit extractor for Web tables, but in scientific publications, the diversity of unit expressions and the use of derived units make unit extraction even harder.  For example, $\text{mag }\text{arcsec}^{-2}$ is a common unit in astrophysics that can also be written as $\text{mag}\cdot \text{arcsec}^{-2}$, $\text{mag}/\text{arcsec}^{2}$, etc.  The research reported here investigates the latter solution -- identifying the target quantity type from the query text.  Once the target quantity is identified, query expansion, which is well understood, can be used to attach expected units to the original query.

Quantity expansion is the task of mapping abstract query terms to quantity descriptors that might be observed in tables.
Quantity types are defined by QUDT\footnote{\url{http://www.qudt.org}}, a quantity type, units, and data type ontology for science and engineering.  QUDT also provides the units commonly used to measure each quantity type.  For example, \texttt{energy} is a quantity type, and electron volt (eV) and Joule (J) are two of many units for \texttt{energy} in QUDT.

Queries such as \texttt{earth-like planet} don't mention a desired quantity type.  Queries such as \texttt{electrical conductivity of materials at different temperatures} mention multiple quantities (electrical conductivity, temperature).  A quantity expansion technique must handle queries that have zero, one, or several quantities.

We assume that if a query refers to a desired quantity type, the type is associated with a query concept (Section \ref{sec:query_concept}).  For example, the query \texttt{gravitational forces in newtonian gravity versus bimetric gravity} (Table \ref{tab:query_example}) is decomposed into three entity concepts (gravitational force, newtonian gravity, versus) and three noun phrase concepts (gravitational force, newtonian gravity, bimetric gravity).
A logistic regression classifier is used to identify the quantity type associated with each concept $c_i$.  
Once the quantity type is determined for query concept $c_i$, its expansion terms are obtained from QUDT by lookup.

During classification, each concept $c_i$ is represented by a vector of four features, described below.

\begin{table}[tbh]
    \centering
    \begin{tabular}{m{3cm}|m{10cm}} \hline
        Query & gravitational forces in newtonian gravity versus bimetric gravity \\ \hline
        Desired Concepts & gravitational force, newtonian gravity, bimetric gravity \\ \hline
        Entities & gravitational force, newtonian gravity, versus \\ \hline
        Noun Phrases & gravitational force, newtonian gravity, bimetric gravity \\ \hline
        Target Quantity & Force, Acceleration \\ \hline
    \end{tabular}
    \caption{Desired concepts, concepts extracted by noun phrases and entity recognition, and target quantity type of an example query.}
    \label{tab:query_example}
\end{table}

\textbf{Language model features:}  A language model for quantity type $u$ (e.g., \texttt{Energy}) is formed from its corresponding Wikipedia page and its common units of measurement from QUDT (e.g., \texttt{keV}).  A background language model that covers dimensionless concepts (e.g., counting numbers) is formed by randomly sampling terms from  Wikipedia.  Each language model is indexed as a pseudo-document by a conventional full-text search engine (e.g., Indri).  

Query concept $c$ is submitted to the search engine to produce a ranking of candidate quantity types $u$ (including the dimensionless type).  The rank and score of the candidate quantity type define two features.
(2 features)

\textbf{Cooccurrence feature:} For each term $w$ of a concept $c_i$, we count the number of times it cooccurred with a unit of quantity type $u$ in the content of publications, and average across all $w \in c_i$ as a feature.  (1 feature)

\textbf{Name overlap}: A Boolean feature indicates whether the name of the quantity or any of its unit symbols appear in the concept.  For example, the concept \texttt{gravitational forces} explicitly uses the term ``forces'', so this feature would be true for the \texttt{Force} quantity type classifier.  (1 feature)

We follow the example of Sarawagi et\ al.~\cite{Sarawagi2014KDD} to generate training data.  They used a high precision unit tagger to automatically collect training data.  In this work, a regular expression program is used to collect term sequence and target quantity type with 97\% Precision.  For example, the unit tagger would extract the pair (\texttt{distance, Length}) from the string \texttt{distance in cm}.
We keep only International System of Units (SI) base types and types derived from SI base types.  Quantity types that do not appear in the corpus are discarded.

A one one-against-all logistic regression classifier was trained for each quantity type.  The prediction is determined by the most confident classifier.  If no classifier gave confidence higher than a threshold of 0.65, the concept was labeled as \texttt{dimensionless}.

To assess the accuracy of the classifier at labeling query concepts, we manually annotated the quantity types of every concept that appeared in a testing query.  Excluding the \texttt{dimensionless} type, which has no effect on the retrieval strategy and also dominates the test set, the Precision of the multi-label classifier is 80.64\%.  Both the threshold and precision were obtained by 5-fold cross validation.

Once a quantity type $u_i$ is identified for each concept $c_i$, we assume a uniform distribution for $p(u_i|q)$ without other prior knowledge.  Each type $u_i$ is represented by joining its units symbols with equal importance.  The query example in Figure~\ref{fig:challenge} is expanded by electron volt with different multipliers, e.g. eV and J, and other common units of energy.

\section{Formulation}
\label{sec:model}
This section shows how to construct structured queries using table fields, key concepts, and quantity units.  We use the query \texttt{forces in newtonian gravity} as a running example.  This query has two concepts:  i) \texttt{forces}, which is associated with a \texttt{force} quantity type; and ii) \texttt{newtonian gravity}, which is associated with an \texttt{acceleration} quantity type.
To simplify the presentation, instead of using all 8 field types, we show only 2 abstract field types, \texttt{field\_1} and \texttt{field\_2}, to represent tables; and we show only two common units for each quantity type: \texttt{N} (newton) and \texttt{lb} (pound) for \texttt{force}, and $m/s^2$ (meters per second) and $km/s^2$ (kilometers per second) for \texttt{acceleration}.

We begin with a brief introduction to the Indri query language, which is used to build structured queries.
\texttt{\#wand} and \texttt{\#and} are weighted and unweighted probabilistic AND query operators, \texttt{\#wsum} is a weighted SUM query operator, and \texttt{\#max} is a operator that selects the best subquery.  \texttt{\#1} matches ngrams, and \texttt{\#uw8} matches when its arguments occur (in any order) within a window of 8 terms.  The suffix \texttt{.(f)} restricts the match to terms that occur within field \texttt{f}.  Metzler and Croft~\cite{MetzlerC04} provide more detail about the Indri query language.

If each table is treated as a document, Indri ranks each table by estimating the probability of table $t$ given query $q$ as shown below~\cite{MetzlerC04}.
\begin{equation}
    \label{eq:init}
    p(t|q) \propto p(t)p(q|t)
\end{equation}
The prior $p(t)$ is estimated by the table quality metric (Section \ref{sect:TableQuality}). 
The structure of query $q$ determines how $p(q|t)$ is estimated.  If $q$ is a bag of words Indri query for the example in Figure \ref{fig:challenge}, as shown below,
\begin{alltt}
\(\texttt{query}\sb{\texttt{BOW}}=\) #and( forces in newtonian gravity )
\end{alltt}
then $p(q|t)$ is estimated as:
\begin{equation}
\label{eq:query_field}
    p(q|t) = \prod_{w \in q} p(w|t)^\frac{1}{|q|}
\end{equation}
where table $t$ is treated as a full-text document with a single bag of words that is the union of information from all field types.  

Earlier studies~\cite{Liu2007AAAI,Pimplikar2012CoRR,Pyreddy1997DL} showed that the roles of table components vary, and thus different table components should be treated differently.  Therefore, following prior work on the use of multiple representations for document retrieval~\cite{OgilvieC03}, we decompose table $t$ into its fields, and estimate $p(q|t)$ as the weighted average of probabilities estimated from each field:
\begin{equation}
\label{eq:query_field}
    p(q|t) = \prod_{w \in q} p(w|t)^\frac{1}{|q|} = \prod_{w \in q} \left( \sum_{f \in t} p(w|f)p(f|t) \right)^\frac{1}{|q|}
\end{equation}
where each $w \in q$ is a word in the query, each $f \in t$ is one of the table fields, and $p(f|t)$ represents the importance of field $f$ to table $t$.
$p(w|f)$ is obtained by estimating a multinomial distribution for $f$ with a maximum likelihood estimator and the well-known two-stage smoothing~\cite{Zhai2002SIGIR}
\begin{equation}
\label{eq:smooth}
    p(w|f) = \lambda\frac{c(w,f)+\mu p(w|\mathcal{T})}{|f|+\mu}+(1-\lambda) p(w|\mathcal{T})
\end{equation}
where $c(w, f)$ is the number of times $w$ appears in field $f$, $|f|$ is the total number of word occurrences in field $f$, $p(w|\mathcal{T})$ is the probability of of $w$ given the whole corpus, and $\lambda, \mu$ are the parameters for Jelinek-Mercer smoothing and Dirichlet prior smoothing respectively.

The structured query below implements Equation~\ref{eq:query_field} for (the abbreviated form of) the example in Table \ref{tab:query_example}.    Different fields are treated as different representations of the table, and the score of each query term is the weighted summation of its score against each field.  
\begin{alltt}
                    \(\texttt{query}\sb{\texttt{terms}}=\) #and( 
                        #wsum( p(field_1|\textit{t}) forces.(field_1)
                               p(field_2|\textit{t}) forces.(field_2) )
                        #wsum( p(field_1|\textit{t}) newtonian.(field_1)
                               p(field_2|\textit{t}) newtonian.(field_2) )
                        #wsum( p(field_1|\textit{t}) gravity.(field_1)
                               p(field_2|\textit{t}) gravity.(field_2) ) )
\end{alltt}

The probabilities \texttt{p(field\_$i$|$t$)} indicate how well field $i$ represents the content of table $t$.  Often in language modeling research, this type of probability is treated as a parameter that is tuned for a dataset; in other words, it represents how well on average field $i$ represents tables in this dataset.

Our goal is a structured query that considers not just query terms, but also the key concepts and quantity units that were recognized in the query (Section~\ref{sec:query}).  We consider these to be conditionally-independent expressions of the underlying information need, thus they are combined by:
\begin{align}
\label{eq:all}
p(q|t) &= p(q_{terms}, q_{concepts}, q_{units}|t) \nonumber \\
&= p(q_{terms}|t)^{1-\alpha-\beta} \times p(q_{concepts}|t)^{\alpha} \times p(q_{units}|t)^{\beta}
\end{align}
where parameters $\alpha$ and $\beta$ control the influence of each component, and $p(q_{terms}|t)$ corresponds to $\texttt{query}_\texttt{terms}$, described above.

$p(q_{concepts}|t)$ represents how well a table matches query concepts.  A query may contain several concepts, and those concepts may have differing importance to the query.  If tables that match all query concepts are to be preferred, then:
\begin{equation}
\label{eq:concept}
p(q_{concepts}|t)=\prod_{c_i \in q}p(c_i|t)^{p(c_i|q)}
\end{equation}
where
$p(c_i|q)$ indicates the importance of concept $c_i$ to query $q$ (Section~\ref{sec:query_concept}).  
The structured query that implements Equation~\ref{eq:concept} for the example query in Table~\ref{tab:query_example} is shown below.
\begin{alltt}
                    \(\texttt{query}\sb{\texttt{concepts}}=\) #wand( 
                        p( \texttt{forces} | q)            \(\texttt{query}\sb{\texttt{c.forces}} \)
                        p( \texttt{newtonian gravity} | q) \(\texttt{query}\sb{\texttt{c.newtonian gravity}}\) )
\end{alltt}

A similar approach is used for $p(q_{units}|t)$, which represents how well a table matches quantitities mentioned by the query.
\begin{equation}
\label{eq:type}
p(q_{units}|t)=\prod_{u_i \in q}p(u_i|t)^{p(u_i|q)}
\end{equation}
$p(u_i|q)$ represents the importance of quantity $u_i$ to query $q$.  As discussed in Section~\ref{sec:query_quantity}, we use uniform weights in this work, although the model permits other types of weights.
The structured query that implements Equation~\ref{eq:type} for the example query is shown below. 
\begin{alltt}
                         \(\texttt{query}\sb{\texttt{quantities}}=\) #wand(
                            p(force|q)        \(\texttt{query}\sb\texttt{u.force}\)
                            p(acceleration|q) \(\texttt{query}\sb{\texttt{u.acceleration}}\)
                         )
\end{alltt}

So far we have incorporated concepts and quantity types at a high level, but have not yet shown how to match them against the table structure.  Recall the visual structure of tables.  They are essentially a kind of data visualization with table headers as axes and cells as values.  Common tables are usually 1-dimension (with only row or column headers) or 2-dimension (with both row and column headers), but there can be more dimensions with tricks such as spanning columns.  As a natural consequence of the visual structure, authors tend to put a concept in one place in the table (e.g., in a row, or in a column) in order to precisely describe the dimensions of a value.  Therefore when we match a key concept or a quantity unit against a table, we select only one field instead of a combination of many fields, hoping that the selected fields can collectively describe the relation between query intents and table structures.  This is an important difference between table retrieval and typical document retrieval using multiple representations.

For each query concept, we use a sequential dependency model (SDM) representation~\cite{Metzler2007SIGIR}.  The sequential dependency model matches query terms that occur anywhere in a table, only in bigrams, and only in short text windows, thus providing a good combination of recall and precision.
In the example below, sequential dependency models match the concept $\texttt{newtonian gravity}$ against each field, and the \texttt{\#max} query operator selects the table field that best represents the concept.
\begin{alltt}
                    \(\texttt{query}\sb{\texttt{c.newtonian gravity}}=\) #max( 
                        #wand( 
                            0.85 #and( newtonian gravity ).(field_1)
                            0.10 #1( newtonian gravity ).(field_1)
                            0.05 #uw8( newtonian gravity ).(field_1) )
                        #wand( 
                            0.85 #and( newtonian gravity ).(field_2)
                            0.10 #1( newtonian gravity ).(field_2)
                            0.05 #uw8( newtonian gravity ).(field_2) )
                    )
\end{alltt}

Each quantity is represented by its common units of measurement, which are obtained from QUDT.  Most quantity units are single terms (e.g., \texttt{N} and \texttt{lb}), for which sequential dependency models have no effect.  Thus, we use a bag of words representation for quantity types and select the field that has the highest generation probability.
\begin{align}
\label{eq:concept_field}
    p(u_i|t) = \max_{f \in t} p(u_i|f)= \max_{f \in t} \prod_{w \in u_i}p(w|f)^\frac{1}{|q|}
\end{align}

Indri's \texttt{\#max} query operator is used to implements Equation~\ref{eq:concept_field} for quantity types, as shown below for the \texttt{force} and \texttt{acceleration} quantity types.
\begin{alltt}
                         \(\texttt{query}\sb{\texttt{u.force}}=\) #max(       
                            #and(N lb).(field_1)         
                            #and(N lb).(field_2)         
                         )                    
                         \(\texttt{query}\sb{\texttt{u.acceleration}}=\) #max(
                            #and(km/s^2 m/s^2).(field_1)
                            #and(km/s^2 m/s^2).(field_2)
                         )
\end{alltt}

\begin{figure}[h]
    \centering
    \includegraphics[width=8cm]{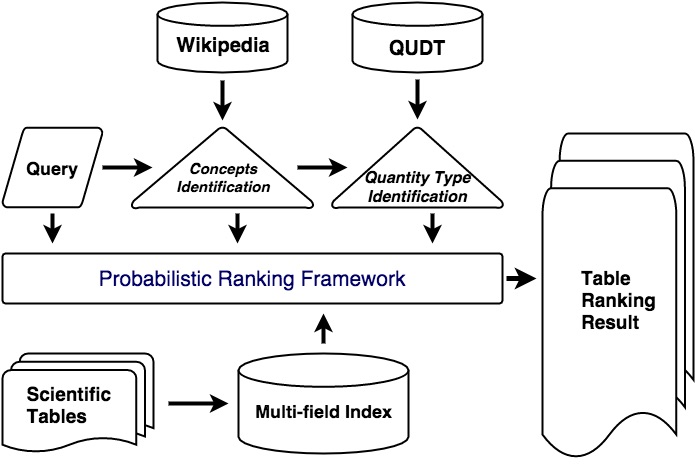}
    \caption{The high-level system architecture.}
    \label{fig:architecture}
\end{figure}

Putting together all three query components, we have an implementation for Equation~\ref{eq:all} that considers terms, concepts, and target quantity types in $q$: 
\begin{alltt}
                            \(\texttt{query}=\) #wand( 
                              \((1-\alpha-\beta)\) \(\texttt{query}\sb{\texttt{terms}}\)
                                      \(\alpha\) \(\texttt{query}\sb{\texttt{concepts}}\)
                                      \(\beta\) \(\texttt{query}\sb{\texttt{quantities}}\)
                            )
\end{alltt}
The model can be interpreted as a combination of term re-weighting (concepts) and query expansion from a controlled vocabulary (quantity types) against a structured document.  Figure~\ref{fig:architecture} shows the high level workflow of the architecture, at the middle of which sits our ranking model.

Since table fields are shorter than traditional documents, the values of smoothing parameters are crucial to ranking performances.  Among different fields, the field length can be as short as a table header, which contains less than 10 words on average, and as long as paper abstracts, whose average length is around 45 words.  For simplicity we use a set of global smoothing parameters for all fields, but in practice it would be natural to have different smoothing parameters for different type of fields.

\section{The T\lowercase{able}A\lowercase{r}X\lowercase{iv} Dataset}
\label{sec:data}
The corpora used in previous table retrieval research were either Web tables or based on documents instead of tables, and the queries were either synthetic~\cite{Liu2007AAAI, Pimplikar2012CoRR} or queries with specific intents~\cite{Sarawagi2014KDD}.  In order to fully investigate the problem of scientific table retrieval and encourage reproducible experiments, we constructed and are making public TableArXiv, a high-quality dataset for scientific table retrieval\footnote{\url{http://www.cs.cmu.edu/~callan/Data/}}.  Below we describe how TableArXiv was created.

arXiv\footnote{\url{http://arXiv.org}} is an e-print repository where scientists can submit draft, working, or published papers.  It accepts papers from multiple scientific domains such as Physics, Computer Science, and Mathematics.  We used a snapshot of arXiv dated Janurary 2014.  The snapshot contains 854,989 papers with metadata, among which Physics and Physics-related domains such as condensed matter and high energy physics comprise 631,734 (73.89\%) papers in total.

We used LaTeXML~\cite{latexml}, a LaTeX processing tool, to convert the LaTeX source files of papers to an XML format, and then extracted 341,573 tables from papers in Physics-related domains.  The average number of tables extracted for each paper is footnotesize because (1) LaTeXML can only convert about 60\% of papers\footnote{\url{http://arxmliv.kwarc.info/}}, and (2) a random sample of 50 papers indicated that about 68\% of Physics papers do not contain tables, while for the remaining 32\% the average number of tables per paper is around 2.67.  

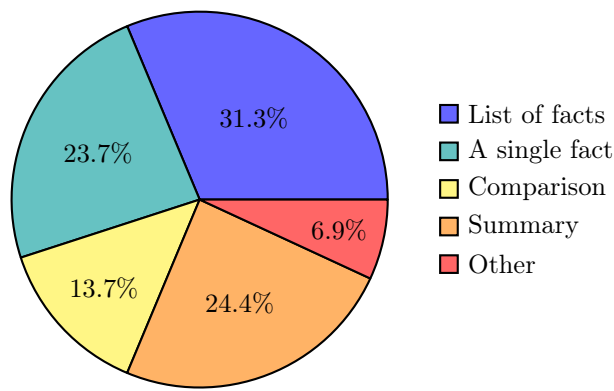
\begin{figure}[h]
    \centering
    \begin{tikzpicture}
    \pie[text=legend, radius=2.5]{31.3/List of facts, 23.7/A single fact, 13.7/Comparison, 24.4/Summary, 6.9/Other}
    \end{tikzpicture}
    \caption{Query Intent Distribution.}
    \label{fig:intent_dist}
\end{figure}

We recruited 8 students (from undergraduates to PhD candidates) majoring in Physics or Physics-related majors to compose information needs in TREC format.
Following recent work on IR evaluation design by Kelly, et al.~\cite{Kelly2015ICTIR}, we asked assessors to assign their information needs to one of five query intent categories.  Table~\ref{tab:query} describes the query intent categories.  Figure~\ref{fig:intent_dist} shows the percentage of queries assigned to each query intent.

For each information need, students assessed a pooled list of at most 100 tables returned by 8 simple ranking algorithms; specifically one bag-of-words ranker for each table field plus another bag-of-words ranker that treated tables as full-text documents.  The order of the result list was randomized to eliminate an ordering bias, and assessors were reminded frequently that the list was in random order.  For each table, assessors chose a rating from a 4-point scale to describe its relevance to his/her information need.  We removed information needs that were duplicated, queries that contained unsupported Unicode characters, and those that had no relevant results according to the assessor.  The result was 105 information needs with relevance judgements, following the Cranfield methodology.

Our 4-point relevance scale was modeled after the relevance values used for TREC 10 \& 11 datasets for Ad-Hoc and Web track.  Table~\ref{tab:rel_dist} compares the the distribution of relevance values for TableArXiv with the distribution of relevance values for TREC 10 \& 11.  The distributions are similar.  

\section{Experimental Methodology}
\label{sec:eval}
This section describes how experiments were conducted.  Search environments are described first, then baseline algorithms, and finally the metrics used for evaluation.

\begin{table}[t]
    \centering
  
    \begin{tabular}{l p{11cm}} \toprule
        \multicolumn{2}{c}{\textbf{Query Intent Categories.}} \\ \midrule
        \bf Intent 1: & \bf A single fact \\ 
        Explanation: & Relevant tables should contain an intended value or fact \\
        Example: & neutron EDM bound \\ \midrule
        \bf Intent 2: & \bf A list of facts \\ 
        Explanation: & Relevant tables should contain a list of related values under different conditions or in relation to different concepts \\
        Example: & speed of light in different media \\ \midrule
        \bf Intent 3: & \bf Comparison \\
        Explanation: & Relevant tables should contain a comparison of values or properties for related concepts \\
        Example: & properties of sun compared with other stars \\ \midrule
        \bf Intent 4: & \bf Summary \\
        Explanation: & Relevant tables should contain general information for the topic \\
        Example: & effect of lensing galaxies on imaging of distant clusters \\ \midrule
        \bf Intent: & \bf Other \\
        Example: & earth-like planet \\ \bottomrule
    \end{tabular}
     \caption{Query intent categories, explanations, and example queries.}
     \label{tab:query}
\end{table}

\begin{table}[t]
    \centering
    \begin{tabular}{c|c|c|c|c|c} \hline
        Data & \#Judged & Non & Rel & HRel & Key \\ \hline
        TREC10 & 23,897 & 78.10\% & 16.81\% & 4.51\% & 0.58\% \\
        TREC11 & 18,362 & 82.81\% & 11.10\% & 3.87\% & 2.22\% \\
        TableArXiv & 9,868 & 83.82\% & 11.22\% & 4.20\% & 0.77\% \\ \hline
    \end{tabular}
    \caption{A comparison of the relevance distributions of judged documents for TableArXiv and TREC 10 \& 11 Ad-Hoc and Web Track datasets.}
    \label{tab:rel_dist}
\end{table}

\subsection{Search Environments and Parameters}
Tables were indexed by the Indri search engine~\cite{MetzlerC04} using its default stopword list and Krovetz stemmer.  Following standard settings in previous work, the top 100 ranked tables were retrieved as results.  The smoothing parameters were determined by 10-fold cross validation.  The Jelinek-Mercer smoothing parameter $\lambda$ was 0.58 when treating tables as full-text documents and 0.81 when using individual field language models.  The Dirichlet prior smoothing parameter $\mu$ was 250 and 2 in the same two conditions.
These values are footnotesizeer than values typically used for document retrieval because tables and table fields are much shorter than Web documents.  All the other parameters, including sequential dependency model weights, the importance of each table field for our structured method and \texttt{BM25F}, and the values of $\alpha$ and $\beta$ in Equation~\ref{eq:all} were also chosen by 10-fold cross validation.  
The partitioning of cross validation was randomized and kept fixed across all experiments.    

\begin{table*}[h]
\centering
\begin{tabular}{l p{11cm}}
Method & Description \\ \toprule
\texttt{Indri-BOW} & Bag-of-words queries. Treat tables as full-text documents. \\ \midrule
\texttt{Indri-SDM} & Sequential dependency model queries.  Treat tables as full-text documents.\\ \midrule
\texttt{BM25} & Bag-of-words queries.  Treat tables as full-text documents.\\ \midrule
\texttt{BM25F} & Bag-of-words queries.  Treat tables as documents that have fields. \\ \midrule
\texttt{TableRank} & Bag-of-words queries.  Treat tables as structured documents. \\ \midrule
\texttt{MaitreD} & The experimental system as presented in Equation~\ref{eq:all}.  It considers table structure, key query concepts, inferred quantity types, and table quality. \\ \midrule
\texttt{MaitreD/Q} & \texttt{MaitreD} except table quality. \\ \midrule
\texttt{MaitreD/QU} & \texttt{MaitreD} except table quality and inferred quantity types\\ \midrule
\texttt{MaitreD/QUC} & \texttt{MaitreD} except table quality, inferred quantity types, and key concepts.  Equivalent to $\texttt{query}_\texttt{terms}$. \\ \bottomrule
\end{tabular}
\caption{Retrieval methods considered in the experiments.}
\label{tab:methods}
\end{table*}

\subsection{Baseline Retrieval Methods}
The experiments were done with five baseline retrieval methods that use varying amounts of query and document structure.  \texttt{BM25} and \texttt{Indri-BOW} use unstructured queries and documents.  \texttt{Indri-SDM} uses structured queries and unstructured documents.  \texttt{BM25F} uses structured queries and documents.  \texttt{TableRank} uses unstructured queries and structured documents.  Among the baselines, only \texttt{TableRank} is dedicated to the table search task, but since it did not perform better than \texttt{BM25F}, the relative comparisons are against \texttt{BM25F} in the experimental analysis.

\texttt{TableRank} is our implementation of an algorithm published by Liu, et al.~\cite{Liu2007AAAI}.  For \texttt{TableRank}, the field weighting  was determined by the same cross validation procedure that was used for \texttt{MaitreD}.  We did not use the query independent (table boost and document boost) information because they were reported to be less important than the TTF-ITTF.

We chose not to include learning-based algorithms such as WebTables~\cite{Cafarella2008PVLDB} as baselines because those methods are supervised whereas our method is unsupervised; it would be surprising if supervised methods were not superior.

\subsection{Metrics} We use mean average precision for the top 100 ranking results (MAP@100) to measure the overall ranking quality, and normalized discounted cumulative gain (NDCG@20) and expected reciprocal rank (ERR@20) for the top 20 results to evaluate the top part of the rankings.  We also show win-tie-loss counts based on MAP@100 to investigate the effects on individual queries.

The best performing results are marked in bold.    Statistical significance is marked by $\dagger$ and $\ddag$ for significant difference ($0.01 \leq p < 0.05$) and highly significant difference ($p<0.01$) respectively using the paired t-test.

\section{Experimental Evaluation}

Four experiments investigated the effectiveness of \texttt{MaitreD}.  The first examined the ranking accuracy of the baseline retrieval methods, \texttt{MaitreD}, and \texttt{MaitreD} with various components disabled.  The second was a more focused comparison of \texttt{MaitreD} and \texttt{TableRank}.  The third experiment compared the effects of entity and noun phrase query concepts.  The last experiment looked more closely at query expansion with units of measurement, and the numeric table quality prior.

\subsection{Ranking Accuracy}
\label{exp:overall}
% Table~\ref{tab:eval_baseline} presents the overall performance of our method versus baselines, and Table~\ref{tab:eval_overall} shows the results of several variants of our model.  \texttt{MaitreD/QUC} is described by Equation~\ref{eq:query_field}.  It uses the original queries and ranks tables based on a weighted combination of scores from different fields.   \texttt{MaitreD/QU} considers concepts according to Equation~\ref{eq:concept}.  The best performance among using different concepts candidates and weighting choices is reported.  \textcolor{red}{\bf I'm not sure what the preceding sentence means.}  \texttt{MaitreD/QUC} further incorporates quantity semantics according to Equation~\ref{eq:all}.  And finally, \texttt{MaitreD} takes into consideration both key concepts and quantity types as well as the numeric proportion table quality prior.  In Table~\ref{tab:eval_baseline} all relative changes in parenthesis are calculated against \texttt{BM25F} \textcolor{red}{\bf (I assume that you mean BM25F)}, while in Table~\ref{tab:eval_overall} relative changes are against \texttt{Maitre/QUC} to show performance gains.

Table~\ref{tab:eval_baseline} presents the ranking accuracy of MaitreD and five baseline algorithms.  As expected, unstructured methods were less effective than structured methods because they mix functionally different table fields.  The three unstructured methods have similar performance, although the Indri bag of words (\texttt{Indri-BOW}) method is slightly better across all metrics.  The results may seem surprising, because sequential dependency models and Okapi BM25 are usually more effective for a wide range of tasks and datasets.  However, table retrieval involves short query concepts and document fields.  A simple application of sequential dependency models is likely to cross concept boundaries in queries and field boundaries in tables, and \texttt{BM25}'s probabilistic model may suffer from skewed distributions across different fields, thus promoting spurious matches.

\begin{table*}[tbh]
\centering
\begin{tabular}{l l c l c l c c}
\bf Method & \multicolumn{2}{c}{\bf NDCG@20} & \multicolumn{2}{c}{\bf ERR@20} & \multicolumn{2}{c}{\bf MAP@100} & \bf Win/Tie/Loss \\ \hline
\texttt{BM25F} & 0.2671 & - & 0.1070 & - & 0.2025 & - & - \\
\texttt{TableRank} & 0.1318$\ddag$ & (-50.66\%) & 0.0603$\ddag$ & (-43.64\%) & 0.0798$\ddag$ & (-60.59\%) & 14/2/89 \\ \hline
\texttt{Indri-BOW} & 0.2523 & (-5.54\%) & 0.1068 & (0.19\%) & 0.1791 & (-11.56\%) & 34/5/66 \\
\texttt{Indri-SDM} & 0.2451 & (-8.24\%) & 0.1019 & (-4.77\%) & 0.1714 & (-15.36\%) & 33/6/66 \\ 
\texttt{BM25} & 0.2383 & (-10.78\%) & 0.0988 & (-7.66\%) & 0.1718$\dagger$ & (-15.16\%) & 34/6/65 \\ \hline
\texttt{MaitreD} & \bf 0.3666$\ddag$ & (37.25\%) & \bf 0.1495$\ddag$ & (39.72\%) & \bf 0.2784$\ddag$ & (37.48\%) & 80/5/20 \\ \hline
\end{tabular}
\caption{Retrieval accuracy of five baseline methods and the experimental system (\texttt{MaitreD}).  $\dagger$ indicates a statistically significant difference compared with \texttt{BM25F} with $0.01 \leq p<0.05$, and $\ddag$ highly statistically significant difference with $p < 0.01$.}
\label{tab:eval_baseline}
\end{table*}

\texttt{BM25F} was the strongest baseline.  It is not surprising that \texttt{BM25F} outperformed the unstructured methods, because it considers table structure and the differing values of table fields.  \texttt{TableRank} performed poorly, which was a surprise.  We defer the discussion of \texttt{TableRank} to Section~\ref{sec:tablerank}.

MaitreD performed significantly better than the strongest baseline, \texttt{BM25F}.  Both systems make use of table structure, and give differing weight to different parts of a table.  However, MaitreD also considers query concepts, implied quantity types, and table quality.  The results were statistically significant ($p<0.01$) across all three metrics.  The win/tie/loss difference is especially noteworthy; 80 out of 105 queries were improved.

Table~\ref{tab:eval_overall} explores how the different components contribute to \texttt{MaitreD}.  The first row (\texttt{MaitreD/QUC}) uses only table structure; it corresponds to $\texttt{query}_\texttt{terms}$ in Section~\ref{sec:model}, and is the language modeling equivalent of \texttt{BM25F}.  It was the weakest of the \texttt{MaitreD} results, although still substantially better than \texttt{BM25F} (Table \ref{tab:eval_baseline}).  Each row adds an additional component:  query concepts (C), inferred quantity types (U), and table quality (Q).  There is an improving trend across all evaluation metrics as additional structural and semantic information is included.  Incorporating query concepts gives a 2--7\% improvement across NDCG, ERR, and MAP.  Considering query concepts and quantity types together produces improvements that are larger (4.2--8.3\%) and statistically significantly better than \texttt{MaitreD} using table structure alone in terms of ERR and MAP.  The numerical table quality prior further improves NDCG, ERR, and MAP by 5.1--10.7\% with statistical significance.  Statistical significance is stronger for NCDG@20 and MAP@100 ($p<0.01$) than for ERR@20 ($p<0.05$).

\begin{table*}[tbh]
\centering
\begin{tabular}{l l c l c l c c}
\bf Method & \multicolumn{2}{c}{\bf NDCG@20} & \multicolumn{2}{c}{\bf ERR@20} & \multicolumn{2}{c}{\bf MAP@100} & \bf Win/Tie/Loss \\ \hline
\texttt{MaitreD/QUC} & 0.3390 & - & 0.1422 & - & 0.2514 & - & - \\
\texttt{MaitreD/QU} & 0.3564$\ddag$ & (5.13\%) & 0.1451 & (2.03\%) & 0.2688$\ddag$ & (6.91\%) & 72/8/25 \\ 
\texttt{MaitreD/Q} & 0.3603$\ddag$ & (6.28\%) & 0.1482$\dagger$ & (4.22\%) & 0.2723$\ddag$ & (8.31\%) & 78/7/20 \\ \hline
\texttt{MaitreD} & \bf 0.3666$\ddag$ & (8.14\%) & \bf 0.1495$\dagger$ & (5.13\%) & \bf 0.2784$\ddag$ & (10.74\%) & 79/8/18 \\ \hline \hline
\end{tabular}
\caption{The contribution of each component to the retrieval accuracy of the experimental system.  Relative change is with respect to a system that uses only table structure.  The statistical significance is marked by $\dagger$ for $0.01 \leq p < 0.05$ and $\ddag$ for $p<0.01$.  System components:  Table quality (Q), inferred quantity types (U), and query concepts (C).  The best performing results are marked in bold.}
\label{tab:eval_overall}
\end{table*}

\subsection{Comparison with TableRank}
\label{sec:tablerank}
In our experiments, \texttt{TableRank} performed worse than unstructured methods (Table~\ref{tab:eval_baseline}).  Admittedly \texttt{TableRank} provides a reasonable way to differentiate table fields by flattening the table structure, however it isn't really designed for queries with multiple terms.  The experimental results suggest that cosine similarity may be dominated by the total number of matched terms instead of overall relevance.  For example, if \texttt{meson mass} were the query, a table that mentions keyword \texttt{mass} many times would be ranked higher than another table that contains each keyword only once.  This problem can be eased by introducing document level relevance signals as done in the original work.  When these are omitted, accuracy suffers.

In order to conduct a fair comparison, we artificially constructed a list of single-term queries and compared \texttt{TableRank} with \texttt{MaitreD/QUC}, our variant that only considers table structure.  The method of collecting single-term queries is as follows.  Recall that for each query, TagMe produced a list of entities together with their confidence scores.  If the highest scoring entity happened to be single-term, we use the term to create a single-term query.  The corresponding relevance judgments remain unchanged.  Take \texttt{meson mass} as an example; both \texttt{meson} and \texttt{mass} are tagged entities but \texttt{meson} has higher confidence score according to TagMe, and \texttt{meson} is single-term, so we used it as a single-term query.  This procedure results in 54 single-term queries.

Table~\ref{tab:eval_tablerank} presents evaluation results on this set of queries.  \texttt{TableRank} outperforms \texttt{Indri-BOW} on all the metrics for single term queries.  However, \texttt{MaitreD/QUC} outperforms both.  \texttt{MaitreD/QUC} is better than \texttt{TableRank} primarily because of the effectiveness of two-stage smoothing whereas \texttt{TableRank} lacks the flexibility to control the influence of field lengths and corpus term frequency.  Note that the relative improvement is less meaningful because of the weak baseline.

\begin{table*}[tbh]
\centering
\begin{tabular}{l l c l c l c c}
\bf Method & \multicolumn{2}{c}{\bf NDCG@20} & \multicolumn{2}{c}{\bf ERR@20} & \multicolumn{2}{c}{\bf MAP@100} & \bf Win/Tie/Loss \\ \hline
\texttt{Indri-BOW} & 0.0867 & - & 0.0285 & - & 0.0710 & - & N/A \\ \hline
\texttt{TableRank} & 0.1019 & (17.53\%) & 0.0288 & (1.05\%) & 0.0762 & (7.32\%) & 29/15/10 \\ 
\texttt{MaitreD/QUC} & \bf 0.1183$\dagger$ & (36.45\%) & \bf 0.0365 & (28.07\%) & \bf 0.0938$\dagger$ & (32.11\%) & 31/14/9 \\ \hline
\end{tabular}
\caption{Retrieval accuracy of two baseline methods and the experimental system using single-term queries.  The best performing results are marked in bold.  (54 single-term queries.)}
\label{tab:eval_tablerank}
\end{table*}

\begin{table*}
\centering
\begin{tabular}{p{1.5cm} l c c c c c c c}
\bf Method &                       & \multicolumn{2}{c}{\bf NDCG@20} & \multicolumn{2}{c}{\bf ERR@20} & \multicolumn{2}{c}{\bf MAP@100} & \bf Win/Tie/Loss \\ \hline
\texttt{MaitreD}                   & \texttt{/QUC} & 0.3390 & - & 0.1422 & - & 0.2514 & - & - \\ \hline
$\texttt{MaitreD}_\texttt{NP}$     & \texttt{/QU} & 0.3451$\ddag$ & (1.80\%) & 0.1427 & (0.35\%) & 0.2619$\ddag$ & (4.18\%) & 72/8/25 \\
$\texttt{MaitreD}_\texttt{NP,wt}$  & \texttt{/QU}& 0.3539$\ddag$ & (4.40\%) & 0.1434 & (0.84\%) & 0.2659$\ddag$ & (5.77\%) & 71/9/25 \\ \hline
$\texttt{MaitreD}_\texttt{Ent}$    & \texttt{/QU}& 0.3477$\ddag$ & (2.57\%) & 0.1450 & (1.97\%) & 0.2626$\ddag$ & (4.46\%) & 70/8/27  \\
$\texttt{MaitreD}_\texttt{Ent,wt}$ & \texttt{/QU}& 0.3564$\ddag$ & (5.13\%) & 0.1451 & (2.03\%) & 0.2688$\ddag$ & (6.91\%) & 72/8/25 \\  \hline
\end{tabular}
\caption{Retrieval accuracy using different methods of identifying and weighting query concepts.  The best performing results are marked in bold.}
\label{tab:concept}
\end{table*}

\subsection{Different Types of Key Concepts}
\label{sec:exp_concept}

Section \ref{sec:query_concept} discussed two approaches to recognizing query concepts:  noun phrases and entities.  The results described in the preceding sections were obtained using entity concepts.  The next experiment investigated the difference between using part of speech tagging and named entity tagging to identify query concepts.  

The baseline for this experiment was \texttt{MaitreD/QUC}, which uses only table structure.  $\texttt{MaitreD}_\texttt{NP}\texttt{/QU}$ used unweighted noun phrase concepts, and $\texttt{MaitreD}_\texttt{NP,wt}\texttt{/QU}$ used noun phrases with normalized estimated confidence scores.  $\texttt{MaitreD}_\texttt{Ent}\texttt{/QU}$ used unweighted entity concepts, and $\texttt{MaitreD}_\texttt{Ent,wt}\texttt{/QU}$ used entity concepts with normalized confidence scores given by TagMe.
% All four of the $\texttt{MaitreD/QU}$ methods are variations of the $\texttt{query}_\texttt{concepts}$ queries presented in Section~\ref{sec:model}.

Experimental results are shown in Table~\ref{tab:concept}.
Using query concepts produced better accuracy than using table structure alone  for both methods, with or without weighting.  The differences were statistically significant ($p<0.01$) for NDCG@20 and MAP@100, but not for ERR@20.
Weighted methods produced higher performance than unweighted methods with respect to all metrics.  Entity concepts produced consistently higher performance than noun phrase concepts for unweighted and weighted versions respectively.

Although entities performed a little better than noun phrases, the results for the two conditions were quite similar.  Further analysis revealed that the two methods of identifying key query concepts had a term overlap of about 90\%; that is, they tended to identify the same concepts.  Entities have the added advantage that they provide the possibility of using linked information in an external resource~\cite{Xiong2015CIKM,Xiong2015ICTIR}, whereas noun phrases do not.  Although we do not explore that possibility in this paper, given the slightly better accuracy provided by entity concepts, we use them as our main method of recognizing query concepts.

\subsection{Query Expansion and the Table Prior}
In Section \ref{exp:overall}, key query concepts improved retrieval accuracy more than query expansion using inferred units of measurement and the numeric table quality prior.  However, these three components affect different numbers of queries.  Query concepts are recognized in nearly all queries, but quantity types are only inferred for some queries.  If the query does not imply a quantity type and units of measurement, there may be no value to promoting tables that contain numeric data.  The last experiment examines this issue.

The set of 105 queries was subdivided based on whether a query did or did not have an inferred quantity type.  49 queries had inferred quantity types and were expanded with units of measurement; 56 queries did not have inferred quantity types and thus could not be expanded.

Note that while this division of the queries allows system behavior to be studied more carefully, it does not necessarily reflect true query intents.  For example, the query \texttt{vapor fraction of water} does not have an inferred quantity type because \texttt{fraction} is a dimensionless concept.  However, a person might infer that this user would prefer tables with numeric content.

Table~\ref{tab:eval_unit_separate} shows the accuracy of several MaitreD variants on the two query sets, as well as AvgGain, the average absolute increase in MAP@100 per winning query, and AvgLoss, the average absolute decrease in MAP@100 per losing query.  The baseline for this experiment is \texttt{MaitreD/QU}, which considers table structure plus key concepts.

\begin{table*}[tbh]
\centering
\begin{tabular}{l l c c c c c c}
 & \bf \footnotesize{Method} & \bf \footnotesize{NDCG@20} & \bf \footnotesize{ERR@20} & \bf \footnotesize{MAP@100} & \bf \footnotesize{Win/Tie/Loss} & \bf \footnotesize{AvgGain} & \bf \footnotesize{AvgLoss} \\ \hline
\multirow{2}{*}{w/ Units (49)} & \texttt{MaitreD/QU} & 0.3255 & 0.1456 & 0.2264 & - & - & - \\ 
                               & \texttt{MaitreD/Q} & 0.3356 & 0.1508 & 0.2346 & 18/2/29 & 0.0489 & 0.0206 \\
                               & \texttt{MaitreD} & 0.3422 & 0.1538 & 0.2397 & 25/2/22 & 0.0509 & 0.0318 \\ \hline
\multirow{2}{*}{w/o Units (56)} & \texttt{MaitreD/QU} & 0.3822 & 0.1443 & 0.3046 & - & - & - \\ 
                                & \texttt{MaitreD} & 0.3905$\dagger$ & 0.1461$\ddag$ & 0.3141$\dagger$ & 30/8/18 & 0.0336 & 0.0178 \\ \hline
\end{tabular}
\caption{Retrieval accuracy using queries with and without inferred units of measurement.  Relative change is with respect to \texttt{MaitreD/QU} in each query set.  $\dagger$ indicates a statistically significant difference compared with MaitreD/Qu with $0.01 \leq p < 0.05$ and $\ddag$ for $p<0.01$.  AvgGain is the average absolute MAP@100 increase per winning query, and AvgLoss the average absolute MAP@100 decrease per losing query.  (49 queries expanded with units of measurement, 54 queries that could not be expanded.)}
\label{tab:eval_unit_separate}
\end{table*}

For queries with units, although average performance improved, the win/tie/loss and average gain and loss show that expanding with quantity units is high-risk/high-reward.  Query expansion improved just a few more queries than it hurt; however, there was a large absolute gain for queries that improved, and a footnotesizeer absolute loss for queries that deteriorated, thus average accuracy improved.  This outcome is typical for query expansion algorithms~\cite{Collins-ThompsonC07}.

Analysis of the 29 losing queries showed that the inferred quantity type sometimes failed to match the units of measurement used in relevant tables.  For example, the quantity type \texttt{Force} was inferred for the query \texttt{forces in newtonian gravity}, which is correct.  However, in relevant tables, authors might use units of \texttt{Mass} and \texttt{Length} as indirect indicators of the \texttt{Force} of gravity.  The relation between these quantity types is known according to the gravity equation $F=\frac{G m_1 m_2}{r^2}$, where $F$ is force, $G$ is a gravity constant, and $m_1$ and $m_2$ are the masses of two objects.  However, MaitreD's query expansion using common units of measurement from QUDT for inferred quantity types is not sufficient to overcome this type of vocabulary mismatch.

The numeric quality prior was effective for both query sets.  Interestingly, it had a more consistent, and thus statistically significant, impact on queries \textit{without} inferred query types.  This result indicates that people who search for tables in the domain of Physics are usually interested in tables that contain numbers, even when that intent is not apparent from the query.

% The subqueries that represent query concepts are constructed using sequential dependency models (SDM) (e.g., the \(\texttt{query}\sb{\texttt{c.newtonian gravity}}\) example in Section \ref{sec:model}).  The next experiment investigated the value of sequential dependency models to this matching process.

% Table~\ref{tab:concept_sdm} presents the experimental results of the effectiveness of SDM.  For \texttt{MaitreD/QUC}, the SDM is applied with a whole query, while for \texttt{Concept} the SDM is only applied with key concepts, which are shorter than queries.  Generally using SDM improves retrieval performances, and applying SDM with concepts instead of the whole query gives comparable results for NDCG@20 and better performance for ERR@20.  This result is intuitive because a query usually contains multiple concepts so that SDM would match not only sequences for concepts but also sequences that cross the boundaries of concepts.

% \begin{table}[h]
% \centering
% \begin{tabular}{p{2.5cm} c c c c}
% \bf Method &  \multicolumn{2}{c}{\bf NDCG@20} & \multicolumn{2}{c}{\bf ERR@20}  \\ \hline
% \texttt{MaitreD/QUC} & 0.3336 & - & 0.1369 & - \\ \hline
% \texttt{MaitreD/QUC (SDM)} & 0.3456 & (3.60\%) & \bf 0.1421 & (3.80\%) \\ \hline
% \texttt{MaitreD/QU (without SDM)} & \bf 0.3540 & (6.12\%) & 0.1411 & (3.07\%) \\
% \texttt{MaitreD/QU} & \bf 0.3536 & (6.00\%) & 0.1410 & (2.99\%) \\ \hline
% \end{tabular}
% \caption{A comparison of applying SDM in different levels. The best performing results are marked in bold.}
% \label{tab:concept_sdm}
% \end{table}

\section{Conclusion}
\label{sec:conclusion}
This paper presents research on ranking tables for science and engineering papers.  Tables are represented as structured objects that contain different types of information that describe the table and its surrounding context at different levels of detail, and that have query-independent quality estimates.  Unstructured bag of words queries are transformed into structured queries that contain entity or noun phrase concepts and that implicitly indicate quantity types.  Query expansion from a curated ontology bridges the mismatch between implied quantity types and units of measurement likely to be found in tables.  A numeric table quality prior biases retrieval toward tables that contain numeric data.  All of this information is incorporated into a probabilistic retrieval that can be used with the widely-used Indri search engine.
% Experiments show that these interventions improve table retrieval quality substantially and significantly compared to several baseline methods.

Experiments show that these interventions produce substantially and significantly more accurate table rankings than four widely-used document retrieval algorithms and one algorithm designed specifically for tables.  These results are confirmed even for single-term queries, which we believe to be the best-case scenario for \texttt{TableRank}.

Query noun phrases have been studied by much prior research, but have not delivered consistently more effective results than distance-based methods such as sequential dependency models.  Our experiments indicate that noun phrases and entities are effective methods of recognizing query concepts for table retrieval, and can be used in addition to sequential dependency models.  In our experiments, noun phrases and entities tended to identify the same concepts.  Entities were slightly more effective, but the difference was footnotesize.

Like most prior query expansion research, we found query expansion with inferred quantity types and units of measurement to be high-risk/high-reward.  Gains were larger than losses, thus results improved on average.  QUDT delivered good terms, but they could be very general (e.g., \texttt{lb}) or so abbreviated (e.g., \texttt{N}) that they might match inappropriately.  Consistency might be improved by reducing the weight of common terms and making better use of context, for example, seeing evidence that the concept \texttt{Newtons} is present before accepting the abbreviation \texttt{N}.

The numeric table quality prior improved accuracy and consistency in this domain.  This result is not surprising for the Physics domain, but we have not seen it in prior table retrieval research.

The probabilistic model presented here provides a flexible framework for incorporating information from more advanced representations of tables and more advanced representations of the information need.  As better approaches are proposed for identifying important concepts in queries or for query expansion, they can be added easily to improve ranking quality.  

The model contains parameters and weights that must be set.  Our research used parameter sweeps or grid searching with cross-validation to set them, which is standard practice.  In an operational environment with more queries and user interaction, it would be natural to use machine learning to obtain more optimal parameter settings.

Our research also produced TableArXiv, a high-quality table retrieval dataset that contains several hundred thousand tables; 105 information needs with descriptions, queries, and query intents; and relevance judgments on a 4-point scale.  The dataset is available for others to use so that our work can be reproduced and extended.  The query intent information has the potential to support intent-dependent studies of table retrieval, which has not been studied in the past.

\section{Acknowledgments}
This research was supported by National Science Foundation (NSF) grant IIS-1450545.  Any opinions, findings, conclusions, and recommendations expressed in this paper are the authors' and do not necessarily reflect those of the sponsors.

\bibliographystyle{apacite}

\end{document}